# State-Based Automation for Time-Restricted Eating Adherence

Samuel E. Armstrong, MS[1], Aaron D. Mullen, BS[1], J. Matthew Thomas, PhD[1], Dorothy D. Sears, PhD[2], Julie S. Pendergast, PhD[1], Jeffrey Talbert, PhD[1], Cody Bumgardner, PhD[1]
[1]University of Kentucky, Lexington, KY, USA; [2]Arizona State University, Tempe, AZ, USA

**Abstract**

*Developing and enforcing study protocols is a foundational component of medical research. As study complexity for participant interactions increases, translating study protocols to supporting application code becomes challenging. A collaboration exists between the University of Kentucky and Arizona State University to determine the efficacy of time-restricted eating in improving metabolic risk among postmenopausal women. This study utilizes a graph-based approach to monitor and support adherence to a designated schedule, enabling the validation and step-wise audit of participants' statuses to derive dependable conclusions. A texting service, driven by a participant graph, automatically manages interactions and collects data. Participant data is then accessible to the research study team via a website, which enables viewing, management, and exportation. This paper presents a system for automatically managing participants in a time-restricted eating study that eliminates time-consuming interactions with participants.*

**Introduction**

Research studies involving human subjects often require regular interactions that include recording study parameters or assessing participants. Traditionally, these interactions take place by manually calling, texting, or emailing; however, this is cumbersome with large participant groups and can lead to data collection errors. Researchers at the University of Kentucky and Arizona State University are conducting a randomized clinical trial to assess the efficacy of time-restricted eating in decreasing metabolic risk. This trial requires collecting daily calorie intake initiation and end time from participants twice daily. Participants then receive encouraging or informative feedback messages about their success rates depending on whether eating took place during the correct time window. At the beginning of the study, these messages were manually sent by the study team, as there were only a few participants enrolled. However, as participant enrollment increased, manual texting became infeasible. An automated method of interaction and management was required to lighten the workload of the research team and to accommodate numerous participants. Several academic[1,2] and commercial[3,4] applications exist but lack features or incur a usage fee. Thus, an in-house solution we call SmartState was initially created to automate many aspects of this study and then expanded into a feature-rich, open-source data collection application currently available for any research study at no cost.

SmartState uses finite state machines (FSMs) to ensure participants receive correct information at specific times of day, provides an auditable log, and offers a strict format for data collection. Originally, FSMs were developed for the purpose of sequential circuit testing[5]. However, their practical applications have expanded across various domains. For example, FSMs are used in robotics to model behavior[6], smart homes for appliance and energy management[7], and automotive systems for autonomous driving[8]. Because FSMs are mathematical models of computation, they can describe the logic within an application in a convenient way. If distinguishable states and transitions exist for a system and can be logically defined, then an FSM can be created to model that system. Likewise, if it is possible to model a study protocol as an FSM, it is also guaranteed that the protocol can be implemented as code into SmartState. As part of the development process, protocols were verified using state graphs and adjusted for logical compliance where needed. At a high level, participant deviation from a set path is eliminated because a participant must enter "checkpoints" before reaching a set "goal." This prevents skipping steps that are crucial for the study's objectives and maintains a systematic progression through the intended structure.

This paper discusses this study in the context of texting, but it is important to note that many other methods of communication are available. For example, emailing and calling participants are also available using a text- and voice-based AI virtual assistant system. All communication methods are abstracted from the underlying state graph, allowing for any communication method to be used without the need for additional implementations.

**Methods**

The accountability and integrity of researchers and, subsequently, the implementation of a research study is imperative. The research study implementation described here monitors and interacts with participants over a 14-hour overnight fast and a 10-hour daytime calorie consumption period. To effectively respond with the appropriate message or alert study administration of a problem, it is essential to ensure the accuracy of the collected starting and ending fasting times. In the case of manual text responses, if a human error occurs, such as forgetting to send a response or sending a wrong response, a participant might be left confused and provide incorrect data. The same scenario can happen in automated systems that are not controlled by state-based logic. In cases where rule-based logic is employed, the potential for programmatical mistakes or incorrect logic can lead to unintended consequences, system errors, and suboptimal outcomes. Such issues are avoided with SmartState due to its FSM-based approach, which ensures accountability and verification for all actions.

Rule-based programming reflects a system where decisions are made based on a set of predefined rules or conditions. These rules typically consist of if-then statements, such that if certain conditions are met, then specific actions are taken. Compared to rule-based approaches, FSMs follow a predefined structure, which removes the possibility of deviation. FSMs are a class of mathematical models employed to represent computational processes. They are characterized by states, transitions, inputs, and outputs, which collectively define their operational framework. States denote distinct conditions or modes within a system, transitions delineate the movement between these states triggered by inputs, inputs serve as stimuli prompting state changes, and outputs encapsulate the consequential actions or responses. State machine diagrams, such as Figure 1, are commonly utilized as visual aids to articulate the operational logic of FSMs, facilitating an unambiguous comprehension of their functionality. A popular tool for creating such diagrams and machines is Umple[9], which integrates modeling concepts and visualizations to produce code representing an FSM in a variety of programming languages. By leveraging Umple, an FSM can be clearly defined with simple unified modeling language (UML) code, which then is able to produce visualizations of the system's behavior. After visual verification, Umple then generates clean and maintainable code that is used with protocol definitions within SmartState. In our case, the Java programming language was chosen to facilitate interactions between the generated FSM code (also in Java) and all other components in the system.

Researchers or end users may not have the technical expertise to understand the computational components in SmartState. Therefore, a management website containing abstractions of FSM functions for participant management is bundled into our application. In many cases, including this article, there are groups within an overarching research study. For example, each participant may be initially enrolled for one month in one group and then moved to another group for three additional months. Both groups, in this case, would have different FSMs powering them. During the transition from one group assignment to another, a participant must also be transitioned from one FSM to another. This process is a complicated feat. Adding, stopping, or restarting an FSM can have unintentional side effects if not handled correctly. SmartState performs several checks before, during, and after sending messages or collecting data, which ensures precise timing and data integrity. Our website streamlines this process, allowing researchers to reassign participants while ensuring a smooth transition and accurate data collection throughout the study. Distilling these complex functions into simple interactions on the website removes unintended and unnecessary human interaction with core components. This prevents data loss and unsuitable responses to participants during a transition.

Even though FSMs offer verification and protocol adherence, exception and error checking must still take place. With any application, especially in healthcare, the integrity of the system is imperative. Therefore, we have implemented checks for the content of messages from participants, runtime exception handling, and fault recovery.

With text messaging (or any open-ended messaging system), there will certainly be typos or misinterpretations. Therefore, the first and foremost check that takes place is whether the intent of the received message can be determined. Participants have a list of text phrases that are allowed in the study, so, in this case, the system simply checks if one of these phrases is in the message. If a participant sends a nonsensical message, this will trigger a response with a helpful explanation about what is expected. This also includes sanitizing all received messages before any processing is started to prevent unauthorized manipulation of the system. The next check is for ambiguity within the message. For example, a message containing "startcal 7" is ambiguous because we do not know if this starting calorie time is 7 AM or PM. Therefore, we send a response message to the participant, asking them to specify "AM" or "PM" in their message. If a valid message is received, then this message is passed to the FSM for processing and possibly transitioning to a new state.

If an unexpected error occurs during the runtime of the application, subsequent attempts to rectify the issue take place. This involves retrying database queries or reloading the participant's state information depending on the nature of the error or exception. In the event of a fault in the computer hosting SmartState (such as a restart), the participant state is not lost. All timers and current states are saved at regular 15-minute intervals. Therefore, once the application has been restarted, participant information can be reloaded and picked up where they left off. Furthermore, the server hosting the application also receives weekly backups in the event of a catastrophic failure.

Auditable logs are also available on the website to ensure accountability of the system and of the research staff, who may make changes over the course of a study. These logs record any changes in the system, including sending and receiving participant messages and changing groups assignments. This documentation records system faults and study team interactions, which can be viewed for a single participant or collectively for all participants.

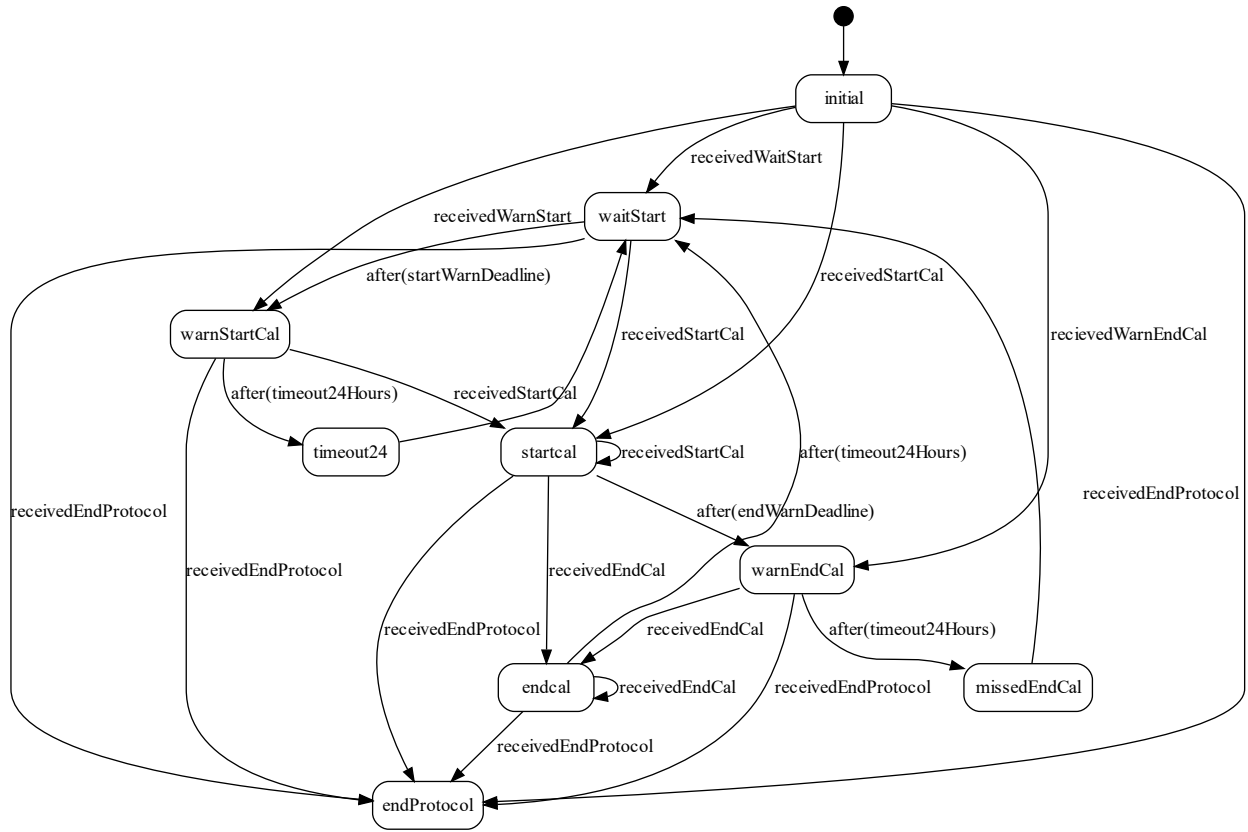

**Figure 1.** Control Group Finite State Machine

The study includes three groups: control, baseline, and time-restricted eating. All participants are initially enrolled in the baseline group. In the baseline group, participants are asked to send a text when they start and end consuming calories for the day. This group does not receive any encouragement or criticism for early or late meals. This allows the study team to measure the duration of eating (time from first to last meal) each day. After two weeks in the baseline group, participants are randomized to either the control group or the time-restricted eating group. The control group continues to report their first and last meals each day, similar to the baseline group. The time-restricted eating group is more complex. Participants in the time-restricted eating group still send start and end messages, and a time window for consuming calories is also calculated. The participants in the time-restricted eating group must consume all their meals in a nine- to eleven-hour window and eat their last meal before 8:00 PM. Thus, a "good" window is between nine and eleven hours from the start of calories to the end of calories. If a participant stops calorie intake too early or too late, they receive informational messages about how to stay within the window. Additionally, a success rate is calculated each day and texted to participants to encourage them to increase their success rate. As this study progressed and became more refined, two additional studies were created and are now also managed by SmartState. Both are similar to the study described in this paper but have key differences, such as different populations and sleep tracking. The study described above, and these two additional studies are managed by the same researchers. Three separate websites for each study were unnecessary as SmartState allows for multiple studies to be encompassed in a single website. Therefore, all three of these studies are available to switch between via a dropdown with each study's name. This consolidated approach streamlines management tasks for researchers while also keeping each study's data independent from the others.

Each group (in any study) is controlled by a unique FSM definition. The FSM representation for the control group is depicted in Figure 1. The control group has many states and transitions. Despite its intricate appearance, this FSM was generated using 107 lines of Umple model code, which generated 619 lines of Java code. This generated code was then integrated into the application in a plug-and-play manner. In our case, we extended this class to perform additional participant management functions, but in many cases, this generated code can be interacted with directly. From the perspective of a researcher, any study requirements can easily be drawn out using this method. The correctness of an FSM can be verified by navigating any paths in the generated diagram. Likewise, from a development standpoint, this approach offers a maintainable means of efficiently translating study requirements into code that can be used with SmartState.

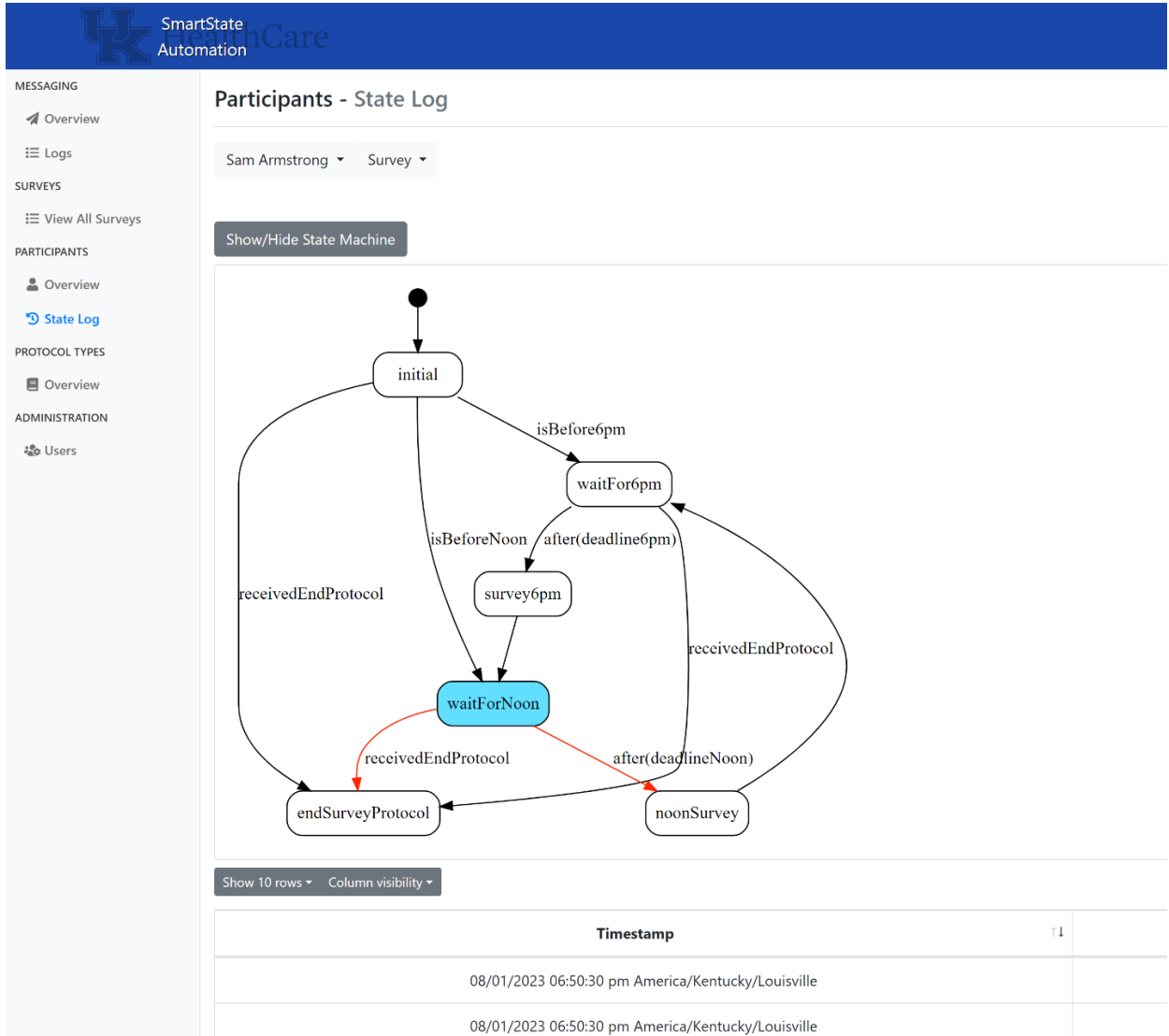

**Figure 2.** Finite State Machine and Audit Log on Web Interface for an Example Protocol

Figure 2 is an example of the web interface and shows an FSM diagram for an active study.
The bridge for communications between participants and the system is Twilio[10], an online-based texting, voice, and video calling service. After purchasing a phone number, requests to send and receive messages on SmartState's behalf can be made. This integration allows for any messages received by the Twilio phone number to also be passed to SmartState via a REST API, which may update a participant's FSM if needed. We chose to use this service over others because the lower cost and the simplicity of code integration into our platform allowed us to quickly begin sending and receiving messages. Once the setup of Umple, Twilio, and the supporting code was complete, a management website was created and connected to the system. The website makes participant management (such as group assignments), viewing messages, exporting collected data, and manually moving states accessible to non-technical users.

Because of the nature of human communication, it is problematic to use a traditional classification system (such as regular expressions) to determine the intent of human-generated phrases or sentences. Since there are many ways to express the same intent, it is necessary to employ artificially intelligent virtual assistants. With the overwhelming success of platforms such as OpenAI's ChatGPT[11], many natural language processing (NLP) tools are free and openly available. Such a tool, namely Rasa[12], is an all-in-one conversational manager, entity extractor, and intent classifier for a loosely defined set of inputs. Comparatively, ChatGPT utilizes many similar techniques for tokenization and entity extraction; however, there is no set response structure for a particular input. Because of the unwieldiness of generative pre-trained transformer (GPT) models, this causes an issue of possible hallucinations or inappropriate responses. The chance of this happening can be reduced by "prompt engineering" to formulating an input in such a way as to increase the chances of a system giving an appropriate response. Prompt engineering is not always reliable because the stochastic nature of GPT models makes it difficult to predict their exact responses. Compared to ChatGPT, Rasa offers restricted, predefined responses for each intent. Especially in healthcare systems, preventing responses that provide incorrect or harmful instructions is essential.

Creating a virtual assistant using Rasa is straightforward, as instructions, examples, and support forums are readily available. However, a drawback of Rasa is the absence of user accessibility, specifically for visually impaired individuals. To create a more robust virtual assistant, an automatic speech recognition (ASR) and text-to-speech (TTS) service is also available in conjunction with Rasa using Nvidia Riva[13]. Riva provides pre-trained ASR models in several languages. If a participant chooses to speak into their device, this audio is transformed into text, which Rasa can parse to generate a textual response. Riva also provides pre-trained TTS models, again, in several languages. The response generated by Rasa can be passed to Riva TTS to produce an audio response in either a male or female voice. This enables voice calls to be carried out automatically with participants without the need for human interaction. To the authors' knowledge, no commercial implementations use both components together in the way described previously. However, by combining these components, participants receive both a streamlined, human-sounding conversation and a customizable chat experience.

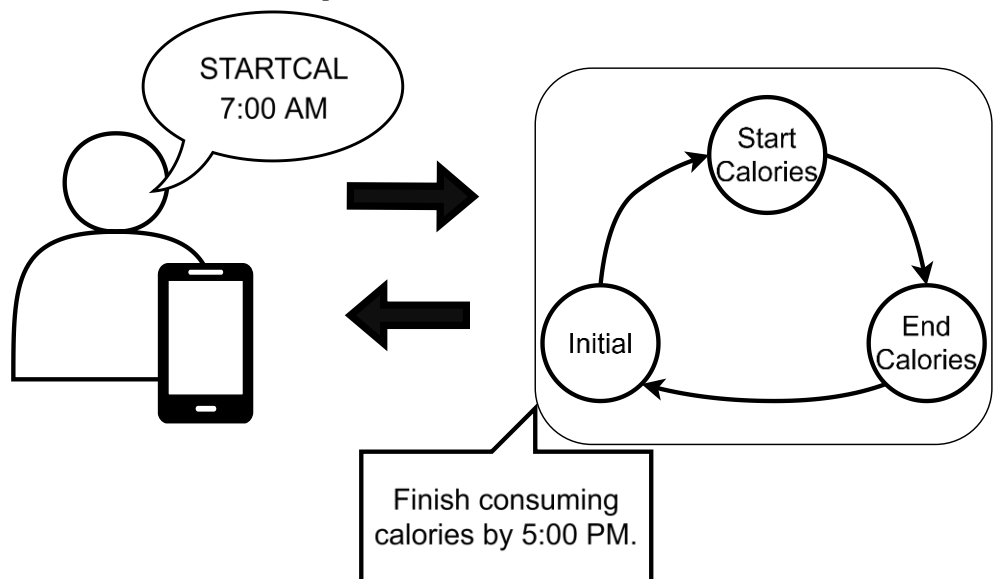

**Figure 3.** Participant texting interaction with state machine

Once a message is received by the system (via text, call, or virtual assistant), it is queued for processing. If the message is well-formatted, then the state machine processes it, transitions to a new state, and sends a response message. Contrary to the full state machine diagram in Figure 1, we show a simplified version of this interaction between a participant and the state machine in Figure 3. In this case, a participant sends the message "STARTCAL 7:00 AM." Initially, the state machine begins in the initial state. Then, once this message is received, the system checks to verify that the message can be parsed. Since this is a valid message, the event to transition to the "Start Calories" state is triggered. Upon entering this state, a 10-hour calorie consumption period is calculated based on the starting time. This calculated time is then sent back to the participant to inform them when they should end their fast. If, in this example, the participant then sent "ENDCAL 4:30 PM," this would again trigger a transition into the "End Calories" state, send a response message, and then wait for the end of the day to automatically transition back to the "Initial" state. At the same time, all well-formatted data is collected and stored for analytical use.

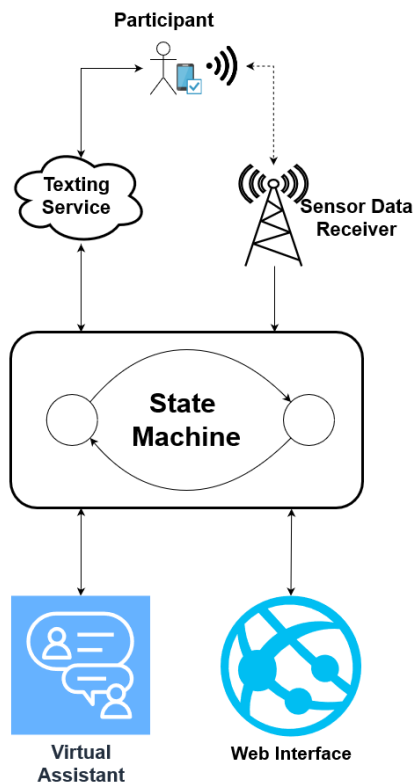

**Figure 4.** Interactions of All Components within the System

Figure 4 shows all components described above and how each interacts within the system. While all components are available to be turned on or off at any time, only a subset of these were necessary to successfully fulfill the system's requirements. With the application's core being the automated FSM interactions, this is the only component that must be enabled for a study to function. Then, specifically for this study, the text messaging service accompanied by the web interface is enabled. While this is a small subset of the capabilities of SmartState, it serves as the optimal configuration for the unique goals of this particular study.

**Analysis**

This automated system aims to eliminate the need for human interactions and to automatically collect start and end times for calorie consumption. Based on a participant's daily calorie consumption window, the success rate is calculated using start and end calorie times. A successful fast is defined by a calorie consumption window of 10 hours (± 1 hour), meaning the fasting time was 14 hours. The number of these successful fasts for a participant is divided by the total number of days a participant is enrolled in the study, which gives them a success rate percentage. Before the creation of SmartState, manual success calculations and interactions were tedious and required a researcher to be ready to respond to a participant as soon as possible. To put this into perspective, a participant must send at least two texts per cycle (starting at 4 AM and ending at 3:59 AM the following day). Then, for each of these messages, a researcher would need to reply to each text with an informational message such as "Keep up the good work" or "You consumed calories for too long." It is essential to monitor these messages in real-time in the event that a participant is fasting for too short or too long in order to prevent unhealthy behavior. Also, if a participant forgets to send a text during the cycle, the researcher will need to send a reminder message. Therefore, a researcher may need to send up to four messages per cycle per participant (startcal reminder, startcal response, endcal reminder, endcal response). Initially, a single person handled all interactions. This researcher estimated it took approximately five minutes per day per participant to send all text messages. Initially, six participants were enrolled in the study, which equated to 30 minutes per day for sending messages and calculating fasting windows.

We additionally measure the efficacy of this system using two metrics: average success rate and average error rate.

The success rate is defined as the average percentage of participants who successfully completed a fast. When enrolled in this group, each participant starts with a success rate of 100%. If a participant consumes calories too long or short, this success rate decreases. To determine the average of this rate, the total number of successful fasts is divided by the total number of days enrolled in the study. This metric only applies to the "Restricted" group. Based on all participants, either currently or previously enrolled in this group, between 9/9/2021 and 3/11/2024, an average success rate of **94%** was achieved.

The error rate is measured by the number of responses sent to participants after the system receives an unknown text message. For example, the most common error occurs when a participant sends an ambiguous time, such as "startcal 7". We do not know whether the participant means 7 AM or PM. Therefore, a response message will be sent saying, "Your STARTCAL time was not understood. Please send 'STARTCAL' again with your starting time including 'am' or 'pm'." This response message would be counted as an error. Other messages that would trigger similar responses include the addition of other words or phrases or misspelling "startcal" or "endcal". From 9/9/2021 to 3/11/2024, 548 messages were not recognized by the system, with the total number of outgoing messages as 5,596. This gives us an error rate of **9.8%.**

After implementing the automatic texting service, no direct messaging occurred between researchers and participants related to the start and end calorie times. Therefore, time spent directly interacting with participants was eliminated. As of March 11th, 2024, the system has received 7,566 messages and sent 5,596 messages automatically without the need for intervention from research staff. When asked if they enjoyed using the automated texting system, participants much preferred the near-instant responses and consistent reminders produced by the system over the manual texts from the research staff.

**Future Work**
We intend to extend our application to include additional data collection methods, specifically continuous health monitoring (CHM) devices. These devices offer a real-time look into a participant's health metrics, and we anticipate that integrating this data would enhance the understanding of participants' progressions within a study. SmartState currently includes implementations for a CHM device network using Wi-Fi, Cellular, and a long-range network called LoRaWAN[14]. By using Wi-Fi and cellular connections, we can achieve a reliable and high bandwidth transmission, making them suitable for situations where real-time updates and high-resolution measurements are paramount. On the other hand, using a LoRaWAN connection enables data collection in remote or challenging environments where traditional connectivity options are limited[15]. At the time of writing, the combination of both short-range, high-bandwidth and long-range, low-bandwidth networks in SmartState have not been tested in a live research study. However, our design is similar to Amazon Sidewalk[16]; therefore, we speculate that a multifaceted CHM device network could reach new populations where they previously were constrained.

**Conclusion**
We have introduced a novel system for tracking and verifying participant protocol adherence throughout a time-restricted eating research study. At its core, SmartState allows for tedious and error-prone data collection to take place programmatically while maintaining an auditable log of participants and system actions at each step of the study.

By utilizing graph-based architecture, specifically finite state machines, we can prevent participants from arbitrarily transitioning between states and, thus, ensure integrity and consistency in their operation. This approach also reduces the workload for developers, enables automatic calculation of participant metrics, decreases message response time, and simplifies participant management for researchers. Abstracting complex studies using Umple and a minimal amount of UML code, we can quickly create FSM implementations in a variety of programming languages to reduce development lead times and prevent unintentional bugs. We built on this FSM base by adding a text messaging system to send and receive participant messages. From these interactions, we calculate statistics such as success rates and fasting timing windows. Our implementation of this automated messaging system outperforms manual methods by increasing the consistency of daily scheduled messages and responding nearly instantly to participant's messages.

Additionally, we offer a website for researchers to manage participants, access audit and message logs, and export participant data for analysis, providing user-friendly functions for interaction with the FSM, enabling researchers to add participants, modify group assignments, monitor current and past participants states, and manually transition participants between states if necessary. We include a virtual assistant for use with the texting system or web-based chat to determine the intent of participant messages, which may vary in grammar, style, and composition. By incorporating speech recognition and generation components into our messaging system, we can also provide a more complete, human-sounding conversation when verbally conversing with a virtual assistant.

The flexibility of our system allows researchers to easily implement their own study requirements according to their specifications. The intermittent fasting study described here shows how SmartState can be applied to automatically manage participant interactions in an efficient, verifiable, and robust way. We have also shown a variety of ways in which this system eliminates manual aspects associated with many research studies to improve the efficiency of data collection.

**Acknowledgement**
Research reported in this publication was supported by the National Institutes of Health grant R01DK124774 and the University of Kentucky Center for Clinical and Translational Science.